\documentclass[a4paper]{jpconf}
\usepackage[T1]{fontenc}
\usepackage[latin9]{inputenc}
\usepackage{geometry}
\usepackage{amssymb}
\usepackage{graphicx}
\usepackage{esint}

\makeatletter
\usepackage{iopams}
\usepackage{setstack}

\usepackage[numbers, sort&compress]{natbib}

\def\newblock{\hskip .11em plus .33em minus .07em}

\makeatother

\begin{document}

\title{{\Large{Hyperbolic groups, 4-manifolds and Quantum Gravity}}}

\author{{\normalsize{Torsten Asselmeyer-Maluga}}}

\address{German Aerospace center, Rosa-Luxemburg-Str. 2, 10178 Berlin, Germany }

\ead{torsten.asselmeyer-maluga@dlr.de}
\begin{abstract}
4-manifolds have special topological properties which can be used
to get a different view on quantum mechanics. One important property
(connected with exotic smoothness) is the natural appearance of 3-manifold
wild embeddings (Alexanders horned sphere) which can be interpreted
as quantum states. This relation can be confirmed by using the Turaev-Drinfeld
quantization procedure. Every part of the wild embedding admits a
hyperbolic geometry uncovering a deep connection between quantum mechanics
and hyperbolic geometry. Then the corresponding symmetry is used to
get a dimensional reduction from 4 to 2 for infinite curvatures. Physical
consequences will be discussed. At the end we will obtain a spacetime
representation of a quantum state of geometry by a non-singular fractal
space (wild embedding) which is stable in the limit of infinite curvatures.
\end{abstract}


%\maketitle

\section{Introduction}

The construction of quantum theories from classical theories, known
as quantization, has a long and difficult history. It starts with
the discovery of quantum mechanics in 1925 and the formalization of
the quantization procedure by Dirac and von Neumann. The construction
of a quantum theory from a given classical one is highly non-trivial
and non-unique. But except for few examples, it is the only way which
will be gone today. From a physical point of view, the world surround
us is the result of an underlying quantum theory of its constituent
parts. So, one would expect that we must understand the transition
from the quantum to the classical world. But we had developed and
tested successfully the classical theories like mechanics or electrodynamics.
Therefore one tried to construct the quantum versions out of classical
theories. In this paper we will go the other way to obtain a quantum
field theory by geometrical methods and to show its equivalence to
a quantization of a classical Poisson algebra. 

The main technical tool will be the noncommutative geometry developed
by Connes \cite{Con:85}. Then intractable space like the leaf space
of a foliation can be described by noncommutative algebras. From the
physical point of view, we have now an interpretation of noncommutative
algebras (used in quantum theory) in a geometrical context. Here we
will use this view to discuss a realization of quantum geometry. Main
idea is the usage of a wild embedding (as induced by an exotic $\mathbb{R}^{4}$),
another expression for a fractal space. Then we will discuss the natural
appearance of a von Neumann algebra (used as observable algebra in
quantum mechanics). A similar relation was found by Etesi \cite{Etesi2017}
based on his previous work \cite{Etesi2016}. Furthermore we will
answer the question whether this algebra is a deformation quantization
of a classical (Poisson) algebra in a positive manner. As a direct
consequence of this correspondence, we will discuss the large (better
infinite) curvature limit of the classical space. This limit agrees
with the corresponding limit for the quantum space. In particular,
we will obtain a dimensional reduction from 4 to 2 for large curvature.
A black hole in this theory admits a non-singular solution with constant
curvature.

\section{From wild embeddings to fractal spaces}

In this section we define wild and tame embeddings and construct a
$C^{*}-$algebra associated to a wild embedding. The example of Alexanders
horned ball is discussed.

\subsection{Wild and tame embeddings\label{sub:Wild-and-tame-embed}}

We call a map $f:N\to M$ between two topological manifolds an embedding
if $N$ and $f(N)\subset M$ are homeomorphic to each other. From
the differential-topological point of view, an embedding is a map
$f:N\to M$ with injective differential on each point (an immersion)
and $N$ is diffeomorphic to $f(N)\subset M$. An embedding $i:N\hookrightarrow M$
is \emph{tame} if $i(N)$ is represented by a finite polyhedron homeomorphic
to $N$. Otherwise we call the embedding \emph{wild}. There are famous
wild embeddings like Alexanders horned sphere or Antoine's necklace.
In physics one uses mostly tame embeddings but as Cannon mentioned
in his overview \cite{Can:78}, one needs wild embeddings to understand
the tame one. As shown by us \cite{AsselmeyerKrol2009}, wild embeddings
are needed to understand exotic smoothness.

\subsection{$C^{*}-$algebras associated to wild embeddings\label{sub:C*-algebra-wild-embedding_idempotent}}

Let $I:K^{n}\to\mathbb{R}^{n+k}$ be a wild embedding of codimension
$k$ with $k=0,1,2$. In the following we assume that the complement
$\mathbb{R}^{n+k}\setminus I(K^{n})$ is non-trivial, i.e. $\pi_{1}(\mathbb{R}^{n+k}\setminus I(K^{n}))=\pi\not=1$.
Now we define the $C^{*}-$algebra $C^{*}(\mathcal{G},\pi$) associated
to the complement $\mathcal{G}=\mathbb{R}^{n+k}\setminus I(K^{n})$
with group $\pi=\pi_{1}(\mathcal{G})$. If $\pi$ is non-trivial then
this group is not finitely generated. The construction of wild embeddings
is usually given by an infinite construction%
\footnote{This infinite construction is necessary to obtain an infinite polyhedron,
the defining property of a wild embedding.%
} (see Antoine\textquoteright{}s necklace or Alexanders horned sphere).
From an abstract point of view, we have a decomposition of $\mathcal{G}$
by an infinite union
\[
\mathcal{G}=\bigcup_{i=0}^{\infty}C_{i}
\]
of level sets $C_{i}$. Then every element $\gamma\in\pi$ lies (up
to homotopy) in a finite union of levels. 

The basic elements of the $C^{*}-$algebra $C^{*}(\mathcal{G},\pi$)
are smooth half-densities with compact supports on $\mathcal{G}$,
$f\in C_{c}^{\infty}(\mathcal{G},\Omega^{1/2})$, where $\Omega_{\gamma}^{1/2}$
for $\gamma\in\pi$ is the one-dimensional complex vector space of
maps from the exterior power $\Lambda^{k}L$ ($\dim L=k$), of the
union of levels $L$ representing $\gamma$, to $\mathbb{C}$ such
that 
\[
\rho(\lambda\nu)=|\lambda|^{1/2}\rho(\nu)\qquad\forall\nu\in\Lambda^{2}L,\lambda\in\mathbb{R}\:.
\]
For $f,g\in C_{c}^{\infty}(\mathcal{G},\Omega^{1/2})$, the convolution
product $f*g$ is given by the equality
\[
(f*g)(\gamma)=\intop_{\gamma_{1}\circ\gamma_{2}=\gamma}f(\gamma_{1})g(\gamma_{2})
\]
with the group operation $\gamma_{1}\circ\gamma_{2}$ in $\pi$. Then
we define via $f^{*}(\gamma)=\overline{f(\gamma^{-1})}$ a $*$operation
making $C_{c}^{\infty}(\mathcal{G},\Omega^{1/2})$ into a $*$algebra.
Each level set $C_{i}$ consists of simple pieces (in case of Alexanders
horned sphere, we will explain it below) denoted by $T$. For these
pieces, one has a natural representation of $C_{c}^{\infty}(\mathcal{G},\Omega^{1/2})$
on the $L^{2}$ space over $T$. Then one defines the representation
\[
(\pi_{x}(f)\xi)(\gamma)=\intop_{\gamma_{1}\circ\gamma_{2}=\gamma}f(\gamma_{1})\xi(\gamma_{2})\qquad\forall\xi\in L^{2}(T),\forall x\in\gamma.
\]
The completion of $C_{c}^{\infty}(\mathcal{G},\Omega^{1/2})$ with
respect to the norm 
\[
||f||=\sup_{x\in\mathcal{G}}||\pi_{x}(f)||
\]
makes it into a $C^{*}$algebra $C_{c}^{\infty}(\mathcal{G},\pi$).
Finally we are able to define the $C^{*}-$algebra associated to the
wild embedding: 

Let $j:K\to S^{n}$ be a wild embedding with $\pi=\pi_{1}(S^{n}\setminus j(K))$
as fundamental group of the complement $M(K,j)=S^{n}\setminus j(K)$.
The $C^{*}-$algebra $C_{c}^{\infty}(K,j)$ associated to the wild
embedding is defined to be $C_{c}^{\infty}(K,j)=C_{c}^{\infty}(\mathcal{G},\pi)$
the $C^{*}-$algebra of the complement $\mathcal{G}=S^{n}\setminus j(K)$
with group $\pi$.

In \cite{AsselmeyerKrol2013} we considered the example of Alexanders
horned ball $A$ as fractal space \cite{Alex:24}. For this example,
the group $\pi_{1}(S^{3}\setminus A)$ is a locally free group of
infinite rank (and perfect). But the last property implies that this
group has the infinite conjugacy class property (icc), i.e. only the
identity element has a finite conjugacy class. This property has a
tremendous impact on the $C^{*}-$algebra \cite{Connes94} and its
enveloping von Neumann algebra: \emph{The enveloping von Neumann algebra
$W(C,\pi_{1}(S^{3}\setminus A))$ of the $C^{*}-$algebra 
\[
C_{c}^{\infty}(C,\pi_{1}(S^{3}\setminus A))
\]
 for the wild embedding $A$ is the hyperfinite factor $I\! I_{1}$
algebra.}

\section{Small exotic $\mathbb{R}^{4}$}

The distinguished feature of differential topology of manifolds in
dimension 4 is the existence of open 4-manifolds carrying a plenty
of non-diffeomorphic smooth structures. In the paper, the special
role is played by the topologically simplest 4-manifold, i.e. $\mathbb{R}^{4}$,
which carries a continuum of infinitely many different smoothness
structures. Each of them except one, the standard $\mathbb{R}^{4}$,
is called \emph{exotic} $R^{4}$. All exotic $R^{4}$ are Riemannian
smooth open 4- manifolds homeomorphic to $\mathbb{R}^{4}$ but non-diffeomorphic
to the standard smooth $\mathbb{R}^{4}$. The standard smoothness
is distinguished by the requirement that the topological product $\mathbb{R}\times\mathbb{R}^{3}$
is a smooth product. There exists only one (up to diffeomorphisms)
smoothing, the standard $\mathbb{R}^{4}$, where the product above
is smooth. In the following, an exotic $\mathbb{R}^{4}$, presumably
small if not stated differently, will be denoted as $R^{4}$.

There are canonical 4-manifolds into which some exotic ${R}^{4}$
are embeddable. Here we will use the defining property of \emph{small}
exotic ${R}^{4}$: every small exotic $R^{4}$ is embeddable in the
the standard $\mathbb{R}^{4}$ (or in $S^{4}$). One of the characterizing
properties of an exotic ${R}^{4}$, which is present in all known
examples, is the existence of a compact subset $K\subset R^{4}$ which
cannot be surrounded by any smoothly embedded 3-sphere (and homology
3-sphere bounding a contractible, smooth 4-manifold), see sec. 9.4
in \cite{GomSti:1999} or \cite{Ganzel2006}. The topology of this
subset $K$ depends strongly on the $R^{4}$. Let $\mathbf{R}^{4}$
be the standard $\mathbb{R}^{4}$ (i.e. $\mathbf{R}^{4}=\mathbb{R}^{3}\times\mathbb{R}$
smoothly) and let $R^{4}$ be a small exotic ${R}^{4}$ with compact
subset $K\subset R^{4}$ which cannot be surrounded by a smoothly
embedded 3-sphere. So, we have the strange situation that an open
subset of the standard $\mathbf{R}^{4}$ represents a small exotic
$R^{4}$.

Now we will describe the construction of this exotic $R^{4}$. Historically
it emerged as a counterexample of the smooth h-cobordism theorem \cite{Don:87,BizGom:96}.
The compact subset $K$ as above is given by a non-canceling 1-/2-handle
pair. Then, the attachment of a Casson handle $CH$ cancels this pair
only topologically. A Casson handle is is a 4-dimensional topological
2-handle constructed by an infinite procedure. In this process one
uses disks with self-intersections (so-called kinky handles) and arrange
them along a tree $T_{CH}$: every vertex of the tree is the kinky
handle and the number of branches in the tree are the number of self-intersections.
Freedman \cite{Fre:82} was able to show that every Casson handle
is topologically the standard open 2-handle $D^{2}\times\mathbb{R}^{2}$.
As the result to attach the Casson handle $CH$ to the subset $K$,
one obtains the topological 4-disk $D^{4}$ with interior $\mathbf{R}^{4}$
o the 1-/2-handle pair was canceled topologically. The 1/2-handle
pair cannot cancel smoothly and a small exotic $R^{4}$ must emerge
after gluing the $CH$. It is represented schematically as $R^{4}=K\cup CH$.
Recall that $R^{4}$ is a small exotic ${R}^{4}$, i.e. $R^{4}$ is
embedded into the standard $\mathbf{R}^{4}$, and the completion $\bar{R}^{4}$
of $R^{4}\subset\mathbf{R}^{4}$ has a boundary given by certain 3-manifold
$Y_{r}$. One can construct $Y_{r}$ directly as the limit $n\to\infty$
of the sequence $\left\{ Y_{n}\right\} $ of some 3-manifolds $Y_{n},n=1,2,...$.
Then the entire sequence of 3-manifolds 
\[
Y_{1}\to Y_{2}\to\cdots\to Y_{\infty}=Y_{r}
\]
characterizes the exotic smoothness structure of $R^{4}$. Every $Y_{n}$
is embedded in $R^{4}$ and into $\mathbf{R^{4}}$. An embedding is
a map $i:Y_{n}\hookrightarrow\mathbf{R^{4}}$ so that $i(Y_{n})$
is diffeomorphic to $Y_{n}$. Usually, the image $i(Y_{n})$ represents
a manifold which is given by a finite number of polyhedra (seen as
triangulation of $Y_{n}$). Such an embedding is tame. In contrast,
the limit of this sequence $n\to\infty$ gives an embedded 3-manifold
$Y_{r}$ which must be covered by an infinite number of polyhedra.
Then, $Y_{r}$ is called a wild embedded 3-manifold (see above). By
the work of Freedman \cite{Fre:82}, every Casson handle is topologically
$D^{2}\times\mathbb{R}^{2}$ (relative to the attaching region) and
therefore $Y_{r}$ must be the boundary of $D^{4}$ (the Casson handle
trivializes $K$ to be $D^{4}$), i.e.\emph{ $Y_{r}$ is a wild embedded
3-sphere $S^{3}$}. $Y_{1}$ was described as the boundary of the
compact subset $K$ whereas $Y_{n}$ is given by $0-$framed surgeries
along $n$th untwisted Whitehead double of the pretzel knot $9_{46}$.
Thus we have a sequence of inclusions 
\[
\ldots\subset Y_{n-1}\subset Y_{n}\subset Y_{n+1}\subset\ldots\subset Y_{\infty}
\]
with the 3-manifold $Y_{\infty}$ as limit. Let $\mathcal{K}_{+}$
be the corresponding (wild) knot, i.e. the $\infty$th untwisted Whitehead
double of the pretzel knot $(-3,3,-3)$ ($9_{46}$ knot in Rolfson
notation). The surgery description of $Y_{\infty}$ induces the decomposition
\begin{equation}
Y_{\infty}=C(\mathcal{K}_{+})\cup\left(D^{2}\times S^{1}\right)\qquad C(\mathcal{K}_{+})=S^{3}\setminus\left(\mathcal{K}_{+}\times D^{2}\right)\label{eq:surgery-description-of-Y}
\end{equation}
where $C(\mathcal{K}_{+})$ is the knot complement of $\mathcal{K}_{+}$.
In \cite{Budney2006}, the splitting of the knot complement was described.
Let $K_{9_{46}}$ be the pretzel knot $(-3,3,-3)$ and let $L_{Wh}$
be the Whitehead link (with two components). Then the complement $C(K_{9_{46}})$
has one torus boundary whereas the complement $C(L_{Wh})$ has two
torus boundaries. Now according to \cite{Budney2006}, one obtains
the splitting 
\[
C(\mathcal{K}_{+})=C(L_{Wh})\cup_{T^{2}}\cdots\cup_{T^{2}}C(L_{Wh})\cup_{T^{2}}C(K_{9_{46}}).
\]
By general arguments (see \cite{AsselmeyerMaluga2016,AsselmeyerKrol2018a})
the complement $C(\mathcal{K}_{+})$ admits a hyperbolic structure,
i.e. it is a homogenous space of constant negative curvature. Therefore
we obtained the first condition: the sequence of 3-manifolds $Y_{1}\to\cdots\to Y_{r}$
is geometrically a sequence of hyperbolic 3-manifolds! By the same
argument, we can also state: \emph{The enveloping von Neumann algebra
$W(Y_{r},\pi_{1}(C(\mathcal{K}_{+}))$ of the $C^{*}-$algebra 
\[
C_{c}^{\infty}(Y_{r},\pi_{1}(C(\mathcal{K}_{+}))
\]
 for the wild embedding 3-sphere is the hyperfinite factor $I\! I_{1}$
algebra.}

\section{Quantum states from wild embeddings}

\begin{figure}
\begin{center}\includegraphics[scale=0.2]{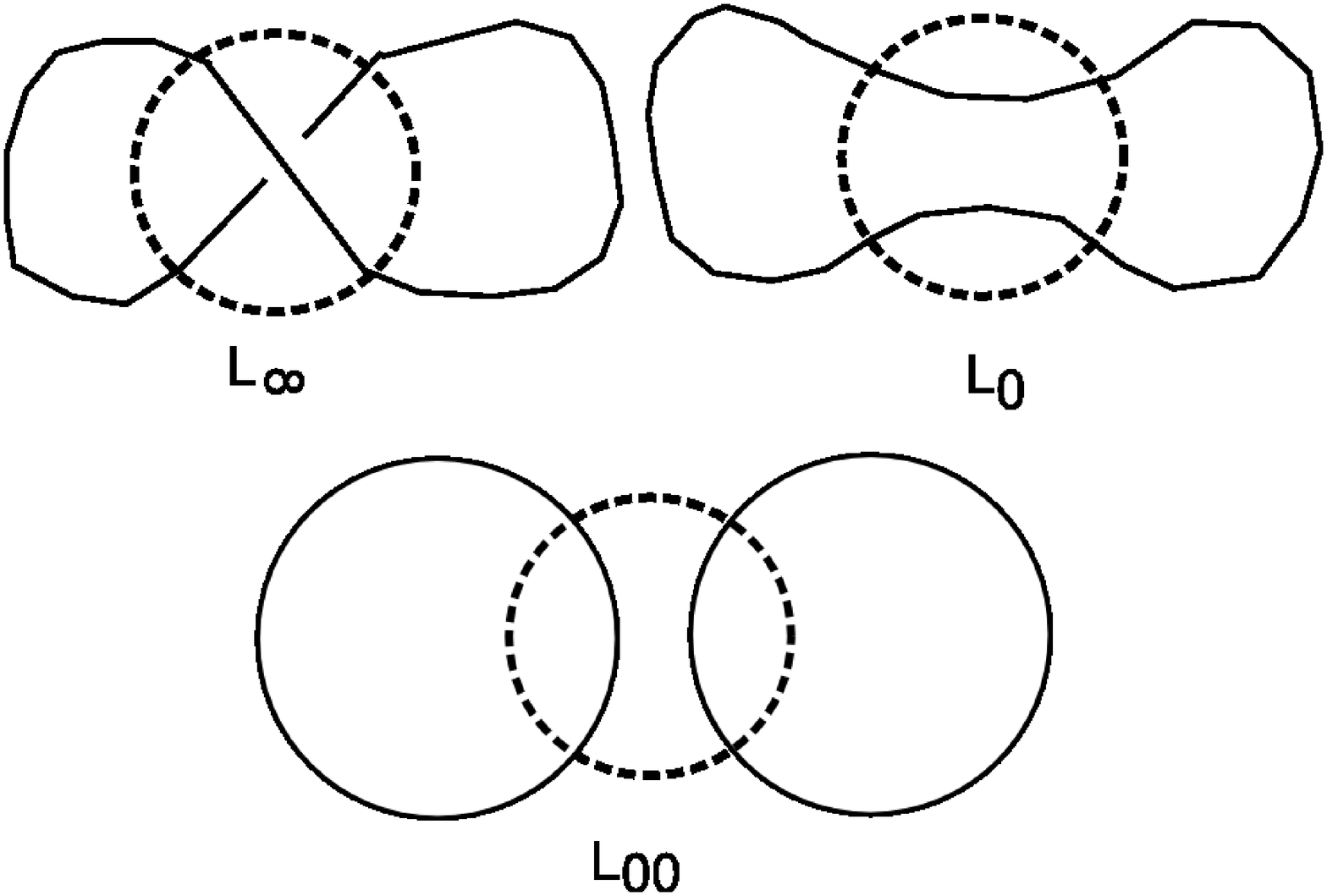}\end{center}

\caption{crossings $L_{\infty},L_{o},L_{oo}$\label{fig:skein-crossings-1}}
\end{figure}

In this section we will describe a way from a (classical) Poisson
algebra to a quantum algebra by using deformation quantization. Therefore
we will obtain a positive answer to the question: Does the $C^{*}-$algebra
of a wild (specific) embedding comes from a (deformation) quantization?
Of course, this question cannot be answered in most generality, i.e.
we use the decomposition of the small exotic $R^{4}$ into the sequence
$Y_{1}\to\cdots\to Y_{r}$. But for this example we will show that
the enveloping von Neumann algebra of this wild embedding (wild 3-sphere
$Y_{r}$) is the result of a deformation quantization using the classical
Poisson algebra (of closed curves) of the tame embedding. This result
shows two things: the wild embedding can be seen as a quantum state
and the classical state is a tame embedding. This result was confirmed
for another case in \cite{AsselmeyerKrol2013} so that we will briefly
list the relevant results (Turaev-Drinfeld quantization):
\begin{itemize}
\item The sequence $Y_{1}\to\cdots\to Y_{r}$ is a sequence of hyperbolic
3-manifolds.
\item The hyperbolic structure is defined by a homomorphism $\pi_{1}(Y_{i})\to SL(2,\mathbb{C})$
($\in Hom(\pi_{1}(Y_{i}),SL(2,\mathbb{C}))$) up to conjugation. 
\item Inside of very $Y_{i}$, there is a special surface $S$ (incompressible
surface) inducing a representation $\pi_{1}(S)\to SL(2,\mathbb{C})$.
\item The space of all representations $X(S,SL(2,\mathbb{C}))=Hom(\pi_{1}(S),SL(2,\mathbb{C}))/SL(2,\mathbb{C})$
has a natural Poisson structure (induced by the bilinear on the group)
and the Poisson algebra \emph{$(X(S,SL(2,\mathbb{C}),\left\{ \,,\,\right\} )$}
of complex functions over them is the algebra of observables. 
\item The Skein module $K_{-1}(S\times[0,1])$ (i.e. $t=-1$) has the structure
of an algebra isomorphic to the Poisson algebra $(X(S,SL(2,\mathbb{C})),\left\{ \,,\,\right\} )$.\emph{
}(see also \cite{BulPrzy:1999,Bullock1999}).
\item The skein algebra $K_{t}(S\times[0,1])$ is the quantization of the
Poisson algebra $(X(S,SL(2,\mathbb{C})),\left\{ \,,\,\right\} )$
with the deformation parameter $t=\exp(h/4)$.(see also \cite{BulPrzy:1999})\emph{
.}
\end{itemize}
To understand these statements we have to introduce the skein module
$K_{t}(M)$ of a 3-manifold $M$ (see \cite{PrasSoss:97}). For that
purpose we consider the set of links $\mathcal{L}(M)$ in $M$ up
to isotopy and construct the vector space $\mathbb{C}\mathcal{L}(M)$
with basis $\mathcal{L}(M)$. Then one can define $\mathbb{C}\mathcal{L}[[t]]$
as ring of formal polynomials having coefficients in $\mathbb{C}\mathcal{L}(M)$.
Now we consider the link diagram of a link, i.e. the projection of
the link to the $\mathbb{R}^{2}$ having the crossings in mind. Choosing
a disk in $\mathbb{R}^{2}$ so that one crossing is inside this disk.
If the three links differ by the three crossings $L_{oo},L_{o},L_{\infty}$
(see figure \ref{fig:skein-crossings-1}) inside of the disk then
these links are skein related. 
Then in $\mathbb{C}\mathcal{L}[[t]]$ one writes the skein relation%
\footnote{The relation depends on the group $SL(2,\mathbb{C})$.%
} $L_{\infty}-tL_{o}-t^{-1}L_{oo}$. Furthermore let $L\sqcup O$ be
the disjoint union of the link with a circle then one writes the framing
relation $L\sqcup O+(t^{2}+t^{-2})L$. Let $S(M)$ be the smallest
submodul of $\mathbb{C}\mathcal{L}[[t]]$ containing both relations,
then we define the Kauffman bracket skein module by $K_{t}(M)=\mathbb{C}\mathcal{L}[[t]]/S(M)$.
We list the following general results about this module:
\begin{itemize}
\item The module $K_{-1}(M)$ for $t=-1$ is a commutative algebra.
\item Let $S$ be a surface then $K_{t}(S\times[0,1])$ caries the structure
of an algebra.
\end{itemize}
The algebra structure of $K_{t}(S\times[0,1])$ can be simple seen
by using the diffeomorphism between the sum $S\times[0,1]\cup_{S}S\times[0,1]$
along $S$ and $S\times[0,1]$. Then the product $ab$ of two elements
$a,b\in K_{t}(S\times[0,1])$ is a link in $S\times[0,1]\cup_{S}S\times[0,1]$
corresponding to a link in $S\times[0,1]$ via the diffeomorphism.
The algebra $K_{t}(S\times[0,1])$ is in general non-commutative for
$t\not=-1$. For the following we will omit the interval $[0,1]$
and denote the skein algebra by $K_{t}(S)$. 

Now we will present the relation between skein spaces and wild embeddings
(in particular to its $C^{*}-$algebra). For that purpose we will
concentrate on the wild embedding $i:S^{3}\to R^{4}$ of $Y_{r}$,
the wild 3-sphere. We will explain now, that the complement $S^{3}\setminus i(D^{2}\times[0,1])$
and its fundamental group $\pi_{1}\left(S^{3}\setminus i(D^{2}\times[0,1])\right)$
can be described by closed curves around tubes (or annulus) $S^{1}\times[0,1]$. 

Let $C$ be the image $C=i(D^{2}\times[0,1])$ decomposed into components
$C_{i}$ so that $C=\cup_{i}C_{i}$. Furthermore, let $C_{i}$ be
the decomposition of $i(D^{2}\times[0,1]$) at $i$th level (i.e.
a union of $D^{2}\times[0,1]$). The complement $S^{3}\setminus C_{i}$
of $C_{i}$ with $n_{i}$ components (i.e. $C_{i}=\sqcup_{1}^{n_{i}}(D^{2}\times[0,1]$)
has the same (isomorphic) fundamental group like $\pi_{1}(\sqcup_{1}^{n_{i}}(S^{1}\times[0,1])$
of $n_{i}$ components of $S^{1}\times[0,1]$. Therefore, instead
of studying the complement we can directly consider the annulus $S^{1}\times[0,1]$
replacing every $D^{2}\times[0,1]$ component. 

Let $C'$ be the boundary of $C$, i.e. in every component we have
to replace every $D^{2}\times[0,1]$ by $S^{1}\times[0,1]$. The skein
space $K_{t}(S^{1}\times[0,1])$ is a polynomial algebra (see the
previous subsection) $\mathbb{C}[\alpha]$ in one generator $\alpha$
(a closed curve around the annulus). Let $TL_{n}$ be the Temperley-Lieb
algebra, i.e. a complex $*-$algebra generated by $\left\{ e_{1},\ldots,e_{n}\right\} $
with the relations
\begin{eqnarray}
e_{i}^{2}=\tau e_{i}\,, & e_{i}e_{j}=e_{j}e_{i}\,:\,|i-j|>1,\nonumber \\
e_{i}e_{i+1}e_{i}=e_{i}\,, & e_{i+1}e_{i}e_{i+1}=e_{i+1}\,,\, e_{i}^{*}=e_{i}\label{Jones-algebra}
\end{eqnarray}
and the real number $\tau$. If $\tau$ is the number $\tau=a_{0}^{2}+a_{0}^{-2}$
with $a_{0}$ a $4n$th root of unity ($a_{0}^{4k}\not=1$ for $k=1,\ldots,n-1$)
then there is an element $f^{(n)}$ with 
\begin{eqnarray*}
f^{(n)}A_{n} & = & A_{n}f^{(n)}=0\\
1_{n}-f^{(n)} & \in & A_{n}\\
f^{(n)}f^{(n)} & = & f^{(n)}
\end{eqnarray*}
in $A_{n}\subset TL_{n}$ (a subalgebra generated of $\left\{ e_{1},\ldots,e_{n}\right\} $
missing the identity $1_{n}$), called the Jones-Wenzl idempotent.
The closure of the element $f^{(n+1)}\in TL_{n+1}$ in $K_{t}(S^{1}\times[0,1]))$
is given by the image of the map $TL_{n+1}\to K_{t}(S^{1}\times[0,1]))$
which maps $f^{(n+1)}$ to some polynomial $S_{n+1}(\alpha)$ in the
generator $\alpha$ of $K_{t}(S^{1}\times[0,1])$). Therefore we obtain
a relation between the generator $\alpha$ and the element $f^{(n)}$
for some $n$. 

The wilderness of $Y_{r}$ is given by a decomposition of $D^{2}\times[0,1]$
into an infinite union of $(D^{2}\times[0,1])-$components $C_{i}$
(in the notation above). But then we have an infinite fundamental
group where every generator is represented by a curve around one $(D^{2}\times[0,1])-$components
$C_{i}$. This decomposition can be represented by a decomposition
of a square (as substitute for $D^{2}$) into (countable) infinite
rectangles. Every closed curve surrounding $C_{i}$ is a pair of opposite
points at the boundary, the starting point of the curve and one passing
point (to identify the component). Every $C_{i}$ gives one pair of
points. Motivated by the discussion above, we consider the skein algebra
$K_{t}(D^{2},2n)$ with $2n$ marked points (representing $n$ components).
This algebra is isomorphic (see \cite{PrasSoss:97}) to the Temperley-Lieb
algebra $TL_{n}$. As Jones \cite{Jon:83} showed: the limit case
$\lim_{n\to\infty}TL_{n}$ (considered as direct limit) is the factor
$I\! I_{1}$. Thus we have constructed the factor $I\! I_{1}$ algebra
as skein algebra.

\emph{Therefore we have shown that the enveloping von Neumann algebra
}
\[
W(C,\pi_{1}(C(\mathcal{K}_{+})))
\]
\emph{ (=the hyperfinite factor $I\! I_{1}$ algebra) is obtained
by deformation quantization of a classical Poisson algebra (the tame
embedding). But then, a wild embedding can be seen as a quantum state.}

\section{Morgan-Shalen compactification and 2D Einstein-Hilbert action}

Above we considered the space $X(Y_{i},SL(2,\mathbb{C})$ of hyperbolic
structures on the 3-manifold $Y_{i}$ now denoted by $\mathcal{M}$
depending on $\pi_{1}=\pi_{1}(Y_{i})$. Let $\rho:\pi_{1}\to SL(2,\mathbb{C})$
be one representation. The character is defined by $\chi_{\rho}(\gamma)=Tr(\rho(\gamma))$
for a $\gamma\in\pi_{1}$. The set of all characters forms an algebraic
variety which is equivalent to $\mathcal{M}$. By the Ambrose-Singer
theorem, the characters (or holonomies in the 3-manifold) are an expression
of the curvature of the 3-manifold. Now we will discuss what happens
for large curvatures or we will discuss the compactification of the
space $\mathcal{M}$. Morgan and Shalen \cite{MorganShalenI1984}
studied a compactification of this space or better they determined
the structure of the divergent signals. The compactification $\overline{\mathcal{M}}$
is defined as follows: let $C$ be the set of conjugacy classes of
$\Gamma=\pi_{1}(\mathcal{N})$, and let $\mathbb{P}(C)=\mathbb{P}(\mathbb{R}^{C})$
be the (real) projective space of non-zero, positive functions on
$C$. Define the map $\vartheta:\mathcal{M}\to\mathbb{P}(C)$ by 
\[
\vartheta(\rho)=\left\{ log(|\chi_{\rho}(\gamma)|+2)\:|\,\gamma\in C\right\} 
\]
and let $\mathcal{M}^{+}$ denote the one point compactification of
$\mathcal{M}$ with the inclusion map $\iota:\mathcal{M}\to\mathcal{M}^{+}$.
Finally, $\overline{\mathcal{M}}$ is defined to be the closure of
the embedded image of $\mathcal{M}$ in $\mathcal{M}\times\mathbb{P}(C)$
by the map $\iota\times\vartheta$. It is proved in \cite{MorganShalenI1984}
that $\mathcal{M}$ is compact and that the boundary points consist
of projective length functions on $\Gamma$ (see below for the definition).
Note that in its definition, $\vartheta(\rho)$ could be replaced
by the function $\left\{ \ell_{\rho}(\gamma)\right\} _{\gamma\in C}$,
where $\ell_{\rho}$ denotes the translation length for the action
of $\rho(\gamma)$ on $\mathbb{H}^{3}$ (3D hyperbolic space) 
\[
\ell_{\rho}(\gamma)=inf\left\{ dist_{\mathbb{H}^{3}}(x,\rho(\gamma)x)\,|\, x\in\mathbb{H}^{3}\right\} 
\]
where $dist_{\mathbb{H}^{3}}$ denotes the (standard) distance in
the 3D hyperbolic space $\mathbb{H}^{3}$. 

Recall that an $\mathbb{R}$-tree is a metric space $(T,d{}_{T})$
such that any two points $x,y\in T$ are connected by a segment $[x,y]$,
i.e. a rectifiable arc isometric to a compact (possibly degenerate)
interval in $\mathbb{R}$ whose length realizes $d_{T}(x,y)$, and
that $[x,y]$ is the unique embedded path from $x$ to $y$. We say
that $x\in T$ is an edge point (resp. vertex ) if $T\setminus\left\{ x\right\} $has
two (resp. more than two) components. A $\Gamma$-tree is an $\mathbb{R}$-tree
with an action of $\Gamma$ by isometries, and it is called minimal
if there is no proper $\Gamma$-invariant subtree. We say that $\Gamma$
fixes an end of $T$ (or more simply, that $T$ has a fixed end) if
there is a ray $R\subset T$ such that for every $\gamma\in\Gamma$
, $\gamma(R)\cap R$ is a subray. Given an $\mathbb{R}$-tree $(T,d_{T})$,
the associated length function $\ell_{T}:\Gamma\to\mathbb{R}^{+}$
is defined by
\[
\ell_{T}(\gamma)=inf_{x\in T}d_{T}(x,\gamma x)
\]
If $\ell_{T}\not=0$, which is equivalent to $\Gamma$ having no fixed
point in $T$ (cf. \cite{MorganShalenI1984,MorganShalenII1988}, Prop.
II.2.15), then the class of $\ell_{T}$ in $\mathbb{P}(C)$ is called
a projective length function.

Now we are able to formulate the main result:

\emph{If $\rho_{k}\in\mathcal{M}$ is an unbounded sequence, then
there exist constants $\lambda_{k}\to\infty$ (renormalization of
the sequence) so that the rescaled length 
\[
\frac{1}{\lambda_{k}}\ell_{\rho_{k}}
\]
converge to $\ell_{\rho_{\infty}}$for $\rho_{\infty}:\Gamma\to Isom(T)$
a representation of $\Gamma$ in the isometry group of the $\mathbb{R}-$tree
$T$, i.e. we have the convergence
\[
\frac{1}{\lambda_{k}}\ell_{\rho_{k}}\Longrightarrow\ell_{T}
\]
}

But what is the meaning of this result? For infinite curvatures, the
underlying 3-manifolds degenerates into a tree which agrees with the
tree of the wild 3-sphere $Y_{r}$. To express it differently, trees
can be seen as hyperbolic spaces of infinite curvature. This result
remained true if we consider the spacetime $Y_{r}\times[0,1]$ with
a Lorentz structure given by a homomorphism $\pi_{1}(Y_{r}\times[0,1])\to SL(2,\mathbb{C})$.
But then we will obtain a dimensional reduction from $3+1$ to $1+1$
with the corresponding reduction of the Einstein-Hilbert action of
the spacetime $S$
\[
\intop_{S}d^{2}x\,\sqrt{-g}e^{-2\phi}\left(R+2(\partial\phi)^{2}+4\lambda^{2}\right)
\]
admitting black hole solutions with no singularity \cite{LemosSa:2D-GRT}.
But how is it related to the quantization via the skein algebra? There
is a relation between the Kauffman bracket skein algebra and lattice
gauge theory \cite{BullockFrohmanKania-Bartoszynska1998}. Again the
curvature is related to the holonomies in the lattice and the hyperbolic
geometry (as defined by $SL(2,\mathbb{C})$) deformed the usual (Euclidean
lattice) to a hyperbolic space. In the limit of infinite curvature
we will obtain a tree again, so meeting our result by using the Morgan-Shalen
compactification. Expressed differently, the large curvature limit
agrees with the quantum description.

\section{Conclusion}

In this paper we discuss a spacetime representation of a quantum geometry
by a fractal space as given by a wild embedding $Y_{r}$. This wild
embedding can be understand as a deformation quantization of a classical
state (Poisson algebra). The state will be formed by equivalence classes
(skein algebra) of knots (as a basis). The whole construction is consistent
with the large curvature limit where the curvature goes to infinity.
For this case the underlying space degenerates to a tree (or the spacetime
is $1+1$ dimensional). By general arguments, this limits agrees with
the corresponding limit for the quantum regime. A black hole in this
theory has no singularity but constant curvature. The details of the
construction will be discussed in forthcoming work.

\ack{}{}

I acknowledge many fruitful discussion with J. Krol.

%\bibliographystyle{BibTeX/iopart-num/iopart-num}
%\addcontentsline{toc}{section}{\refname}\bibliography{diffbib,foliation-gerbes,knots}
\providecommand{\newblock}{}

\end{document}